\documentstyle[prd,aps, epsfig]{revtex}
\begin{document}

\title{{\bf  
The projector on physical states in loop quantum gravity}}
\author{{\bf Carlo Rovelli}\\ \vskip.2cm
{\it Centre de Physique Theorique\\
CNRS-Luminy, Case 907,
Marseille F-13288, France}\\
and \\
{\it Department of Physics and Astronomy \\ 
University of Pittsburgh, Pittsburgh 
Pa 15260, USA}\\
rovelli@cpt.univ-mrs.fr} 
\date{\today}
\maketitle

\begin{abstract}  
We construct the operator that projects on the physical states in loop 
quantum gravity.  To this aim, we consider a diffeomorphism invariant 
functional integral over scalar functions.  The construction defines a 
covariant, Feynman-like, spacetime formalism for quantum gravity and 
relates this theory to the spin foam models.  We also discuss how 
expectation values of physical quantity can be computed.
\end{abstract} 

\section{Introduction}

The loop approach to quantum gravity \cite{rs} based on the 
Ashtekar variables \cite{Ashtekar} has been successful in 
establishing a consistent and physically reasonable framework for 
the mathematical description of quantum spacetime \cite{Review}.  
This framework has provided intriguing results on the quantum 
properties of space, most notably detailed quantitative results 
on the discrete quanta of the geometry \cite{discreteness}.  The 
nonperturbative {\em dynamics\/} of the quantum gravitational 
field, however, is not yet well understood.  Two major questions 
are open.  First, several versions of the hamiltonian constraint 
have been proposed \cite{Thiemann,rr,Marko}, but the physical 
correctness of these versions has been questioned 
\cite{Marolfetal}.  Second, a general scheme for extracting 
physical consequences from a given hamiltonian constraint, and 
for computing expectations values of physical observables is not 
available.  

In this paper, we address the second of these problems -- a 
solution of which, we think, is likely to be a prerequisite for 
addressing the first problem (the choice of the correct 
hamiltonian constraint).  The problem we address is how 
expectation values of physical observables can be computed, given 
a hamiltonian constraint operator $H(x)$.  (For an earlier 
attempt in this direction, see \cite{old}.)  We address this 
problem by constructing the ``projector'' $P$ on the physical Hilbert 
space of the theory, namely on the space of the solutions of the 
hamiltonian constraint equation.  Formally, this projector can be 
written as 
\begin{equation}
	P \sim \prod_{x} \delta(H(x)) 
	\sim \int [DN]\ e^{-i\int d^{3}x\ N(x) H(x)} 
	\label{formal}
\end{equation}
in analogy with the representation of the delta function as the 
integral of an exponential.  The idea treating first class constraints  
in the quantum theory by using a projector operator defined by a 
functional integration has been studied by Klauder \cite{klauder}, and 
consider also by Govaerts \cite{govaerts}, Shabanov and Prokhorov 
\cite{shabanov}, Henneaux and Teitelboim \cite{henneaux} and others.  
In the case of gravity, the matrix elements of $P$ between two states 
concentrated on two 3-geometries $g$ and $g'$ can be loosely 
identified with Hawking's propagator $P(g,g')$, which is formally 
written in terms of a functional integral over 4-geometries 
\cite{Hawking}.

A step towards the definition of the projector $P$ was taken in 
\cite{rr}, where a perturbative expression for the exponential of the 
hamiltonian smeared with a constant function $N(x)=N$ was constructed.  
What was still missing was a suitable diffeomorphism invariant notion 
of functional integration over $N(x)$.  Here, we consider an 
integration on the space of the scalar functions $N(x)$.  This 
integral, modeled on the Ashtekar-Lewandowski construction \cite{AL} 
and considered by Thiemann in the context of the general covariant 
quantization of Higgs fields \cite{Higgs}, allows us to give a meaning 
to the r.h.s.\ of (\ref{formal}).  Using it, we succeed in expressing 
the (regularized) matrix elements of the projector $P$ in a well 
defined power expansion.  We then give a preliminary discussion of the 
expectation values of physical observables.  The construction works 
for a rather generic form of the hamiltonian constraint, which 
includes, as far a we know, the various hamiltonians proposed so far.

As realized in \cite{rr}, the terms of the expansion of $P$ are 
naturally organized in terms of a four-dimensional Feynman-graph-like 
graphic representation.  The expression (\ref{formal}) can thus be 
seen also as the starting point for a {\em spacetime\/} representation 
of quantum gravity.  Here, we complete the translation of canonical 
loop quantum gravity into covariant spacetime form initiated in 
\cite{rr}.  The ``quantum gravity Feynman graphs'' are two-dimensional 
colored branched surfaces, and the theory takes the form of a ``spin 
foam model'' in the sense of Baez \cite{Baez}, or a ``worldsheet 
theory'' in the sense of Reisenberger \cite{Reisenberger}, or a 
``theory of surfaces'' in the sense of Iwasaki \cite{Iwasaki}, and 
turns out to be remarkably similar to the Barret-Crane model 
\cite{BarretCrane} and to the Reisenberger model \cite{Reisenberg97} 
(see also \cite{geometria}).  On the one hand, the construction 
presented here provides a more solid physical grounding for these 
models; on the other hand, it allows us to reinterpret these models as 
proposals for the hamiltonian constraint in quantum gravity, thus 
connecting two of the most promising directions of investigations of 
quantum spacetime \cite{india}.

The paper is organized as follows.  In Section \ref{loop}, the basics 
of loop quantum gravity are reviewed, organized from a novel and 
simpler perspective, which does not require the cumbersome 
introduction of generalized connections, or projective limits (see 
also \cite{primer}).  Section \ref{regularization} presents the 
definition of the diffeomorphism invariant functional integral.  In 
section \ref{dynamics} we construct the projector $P$ and discuss the 
construction of the expectation values.

\section{Loop quantum gravity} \label{loop}

General relativity can be expressed in canonical form in term of 
a (real) $SU(2)$ connection $A$ defined over a 3d manifold 
$\Sigma$ \cite{Ashtekar,Barbero}.  We take $\Sigma$ to be 
topologically $S_{3}$.  The dynamics is specified by the usual 
Yang-Mills constraint, which generates local $SU(2)$ 
transformation, the diffeomorphism constraint $D[\vec f]$, which 
generates diffeomorphisms of $\Sigma$, and the Hamiltonian 
constraint $H[N]$, which generates the evolution of the initial 
data in the (physically unobservable) coordinate time.  Here 
$\vec f$ is in the algebra of the group $Diff$ of the 
diffeomorphisms of $\Sigma$, namely it is a smooth vector field 
on $\Sigma$, and $N$ is a smooth scalar function on $\Sigma$.  
The theory admits a nonperturbative quantization as follows.  
(For a simple introduction, see \cite{primer}, for details see 
\cite{Review} and references therein.) 

\subsection{Hilbert space and spin networks basis}

We start from the linear space $L$ of quantum states $\Psi(A)$ 
which are continuous (in the sup-topology) functions of (smooth) 
connections $A$.  A dense (in $L$'s pointwise topology) subset of 
states in $L$ is formed by the graph-cylindrical states 
\cite{AL}.  A graph-cylindrical state $\Psi_{\Gamma,f}(A)$ is a 
function of the connection of the form
\begin{equation}
\Psi_{\Gamma,f}(A)	= f(U(e_{1},A), \ldots, U(e_{n}A)),
\label{al}
\end{equation}
where $\Gamma$ is a graph embedded in $\Sigma$, $e_{1} \ldots 
e_{n}$ are the links of $\Gamma$, $U(e,A) = {\cal P}exp 
\int_{e}A$ is the parallel propagator $SU(2)$ matrix of $A$ along 
the path $e$, and $f$ is a complex valued (Haar-integrable) 
function on $[SU(2)]^{n}$.  The function $\Psi_{\Gamma,f}(A)$ has 
domain of dependence 
on the graph $\Gamma$; one can always replace $\Gamma$ with a 
larger graph $\Gamma'$ such that $\Gamma$ is a subgraph of $\Gamma'$, 
by simply taking $f$ independent from the group elements corresponding 
to the links in $\Gamma'$ but not in $\Gamma$.  Therefore any two 
given graph-cylindrical functions can always be viewed as defined on 
the same graph $\Gamma$.  Using this, a scalar product is defined on 
any two cylindrical functions by
\begin{equation}
(\Psi_{\Gamma,f}, \Psi_{\Gamma,g}) = \int_{[SU(2)]^{n}} 
dU_{1} \ldots dU_{n}\ \overline{f(U_{1} \ldots U_{n})}\
 g(U_{1} \ldots U_{n}) 
\end{equation}
and extends by linearity and continuity to a well defined 
\cite{AL,Mathfound} scalar product on $L$.  The Hilbert completion of 
$L$ in this scalar product is the Hilbert space $H_{aux}$: the quantum 
state space on which quantum gravity is defined.\footnote{$H_{aux}$ is 
the state space of the old loop representation \cite{rs}, equipped 
with a scalar product which was first obtained through a path 
involving $C^{*}$-algebraic techniques, generalized connections and 
functional measures \cite{AL,Mathfound}.  Later, the same scalar 
product was defined algebraically in \cite{rovellidepietri} directly 
from the old loop representation.  The construction of $H_{aux}$ given 
here is related to the one in \cite{AL,Mathfound} but does not require 
generalized connections, infinite dimensional measures or the other 
fancy mathematical tools that were employed at first.} We refer to 
\cite{primer} for the construction of the elementary quantum field 
operators on this space.  

The $SU(2)$ gauge invariant states form a liner subspace $L_{0}$ in 
$L$.  A convenient orthonormal basis in $L_{0}$ is the spin network 
basis \cite{spinnet}, constructed as follow.  Consider a graph 
$\Gamma$ embedded in $\Sigma$.  To each link $e$ of $\Gamma$, assign a 
nontrivial $SU(2)$ irreducible representation $j_{e}$, which we denote 
the color of the link.  Consider a node $n$ of $\Gamma$, where the 
links $e_{1}\ldots e_{N}$ meet; consider the invariant tensors $v$ on 
the tensor product of the representation $j_{e_{1}}\ldots j_{e_{N}}$ 
of the links that meet at the node; the space of these tensors is 
finite dimensional (or zero dimensional) and carries an invariant 
inner product.  Choose an orthogonal basis in this space\footnote{For 
later convenience, we choose a basis that diagonalizes the volume 
operator \cite{discreteness,volume}.}, and assign to each node $n$ of 
$\Gamma$ one element $v_{n}$ of this basis.  A spin network 
$S=(\Gamma,\{j\},\{v\})$ is given by a graph $\Gamma$ and an 
assignment of a color $j_{e}$ to each link $e$ and a basis invariant 
tensor $v_{n}$ to each node $n$.

The spin network state $\Psi_{S}(A)$ is defined as 
\begin{equation}
\Psi_{S}(A) = \prod_{e} \prod_{n}\ v_{n}\ R^{j_{e}}(U(e,A))
\end{equation} 
where $R^{j}(U)$ is the matrix representing the $SU(2)$ group 
element $U$ in the spin-$j$ irreducible representation, and the 
two matrix indices of $R^{j_{e}}(U(e,A))$ are contracted into the 
two tensors $v_{n}$ of the two nodes adjacent to $e$.  An easy 
computation shows that (with an appropriate normalization of the 
basis states $v_{n}$ \cite{rovellidepietri}) the states 
$\Psi_{S}$ form an orthonormal basis in $H_{aux}$
\begin{equation}
	(\Psi_{S},\Psi_{S'})=\delta_{\Gamma,\Gamma'} 
	\delta_{\{j\}\{j'\}}
	\delta_{\{v\}\{v'\}}. 
\end{equation}

\subsection{Diffeomorphisms}\label{diff}

The Hilbert space $H_{aux}$ carries a natural {\em unitary\/} 
representation ${\cal U}(Diff)$ of the diffeomorphism group of 
$\Sigma$.
\begin{equation}
	[{\cal U}(\phi)\psi](A)=\psi(\phi^{-1}A), \ \ \ \ \ \phi\in Diff.
\end{equation} 
It is precisely the fact that $H_{aux}$ carries this representation 
which makes it of crucial interest for quantum gravity.  In other 
words, $H_{aux}$ and its elementary quantum operators represent a 
solution of the problem of constructing a representation of the 
semidirect product of a Poisson algebra of observables with the 
diffeomorphisms.\cite{Isham}

Notice that ${\cal U}$ sends a state of the spin network basis 
into another basis state
\begin{equation}
    [{\cal U}(\phi)\psi_{S}](A)=\psi_{S}(\phi^{-1}A)= \psi_{\phi S}(A). 
\end{equation}
Intuitively, the space $H_{diff}$ of the solutions of the quantum 
gravity diffeomorphism constraint is formed by the states 
invariant under ${\cal U}$.  However, no finite norm state is 
invariant under ${\cal U}$, and generalized-state techniques are 
needed.  We sketch here the construction of $H_{diff}$ 
\cite{rs,Mathfound,stati}, because the solution of the hamiltonian 
constraint will be given below along similar lines.  $H_{diff}$ 
is defined first as a linear subset of $L^{*}$, the topological 
dual of $L$.  It is then promoted to a Hilbert space by defining 
a suitable scalar product over it.  $H_{diff}$ is the linear 
subset of $L^{*}$ formed by the linear functionals $\rho$ such 
that
\begin{equation} 
\rho({\cal U}(\phi)\psi) = \rho(\psi) 
 \label{rho}
\end{equation}
for any $\phi\in Diff$.  From now on we adopt a bra/ket notation. We
write (\ref{rho}) as
\begin{equation}
\langle \rho|{\cal U}(\phi) \psi\rangle = \langle \rho|\psi\rangle 
\end{equation}
and we write the spin network state $\Psi_{S}$ as $|S\rangle$.  

Equivalence classes of embedded spin networks under the action of Diff 
are denoted as $s$ and called s-knots, or simply spin networks.  We 
denote as $s(S)$ the equivalence class to which $S$ belong.  Every 
s-knot $s$ defines an element $\langle s|$ of $H_{diff}$ via
\begin{eqnarray}
\langle s | S \rangle  &=& 0,\ \ \   \rm{if}\ \  s \neq s(S) 
\nonumber \\
 &=& c_{s},\ \ \  \rm{if}\ \  s = s(S).  
\label{sstate}
\end{eqnarray}
Here $c_{s}$ is the integer number of isomorphisms (including the 
identity) of the (abstract) colored graph of $s$ into itself that 
preserve the coloring and can be obtained from a diffeomorphism of 
$\Sigma$.  A scalar product is then naturally defined in $H_{diff}$ by
\begin{equation}
\langle s | s' \rangle  \equiv \langle s | S' \rangle  
\label{scalardiff}
\end{equation} 
for an arbitrary $S'$ such that $s(S')=s'$.  One sees immediately 
that the normalized states $\frac{1}{\sqrt{c_{s}}}|s\rangle$ form an 
orthonormal basis.  

The space $H_{diff}$ is not a subspace of $H_{aux}$ 
(because diff invariant states have ``infinite norm''). 
Nevertheless, an important observation is that there is 
a natural ``projector'' $\Pi$ from $H_{aux}$ to $H_{diff}$
\begin{equation}
	\Pi: |S\rangle \longmapsto |s(S)\rangle
	\label{Uprojector}
\end{equation} 
which sends the state in $H_{aux}$ associated to an embedded spin 
network $S$ into the state in $H_{diff}$ associated to the 
corresponding abstract spin network state $s$.  Notice that $\Pi$ is 
not really a projector, since $H_{diff}$ is not a subspace of 
$H_{aux}$, but we use the expression ``projector'' nevertheless, 
because of its physical transparency.  Since $H_{diff}$ can be seen as 
a subspace of $H_{aux}^{*}$, the operator $\Pi$ defines a 
(degenerate)  quadratic form $\langle \ | \ \rangle_{diff}$ on $H_{aux}$
\begin{equation}
	\langle S| S'\rangle_{diff} \equiv 
	\langle S|\Pi| S'\rangle = \langle s(S)|S'\rangle
	= \langle s(S)| s(S')\rangle .
	\label{rl}
\end{equation}
$H_{diff}$ is can be defined also by starting with the pre-Hilbert 
space $H_{aux}$ equipped with the degenerate a quadratic form $\langle 
\ | \ \rangle_{diff}$, and factoring and completing the in the Hilbert 
norm defined by $\langle \ |\ \rangle_{diff}$ .\cite{Landsman} That 
is, states in $H_{aux}$ are the (limits of sequences of) equivalence 
classes of states in $H_{aux}$ under $\langle \ |\ \rangle_{diff}$.  
In other words, knowing the ``matrix elements''
\begin{equation}
	\langle S|\Pi| S'\rangle
\end{equation}
of the projector $\Pi$ is equivalent to having solved the 
diffeomorphism constraint.

Furthermore, the above construction can be expressed also in 
terms of certain formal expressions, which are of particular 
interest because they can guide us in solving the hamiltonian 
constraint.  Define a formal integration over the 
diffeomorphism group $Diff$ satisfying the two properties
\begin{equation}
\int_{Diff} D\phi = 1 ,
\label{reg1}
\end{equation}
and
\begin{equation}
\int_{Diff} D\phi\ \delta_{S,\phi S} = c_{s(S)}. 
\label{reg2}
\end{equation}
Then a diff invariant state $|s\rangle$ can be written as a 
``state in $H_{aux}$ integrated over the diffeomorphism group''. 
That is
\begin{equation}
	|s(S)\rangle = \int_{Diff} D\phi \ | {\cal U}(\phi) S\rangle,
\label{diffsmear}
\end{equation}
In fact, the equations (\ref{sstate}) and (\ref{scalardiff}) can 
be obtained from the equations (\ref{reg1}), (\ref{reg2}) and 
(\ref{diffsmear}).  Using this, we can write the 
projection operator $\Pi$, defined in (\ref{Uprojector}), as
\begin{equation}
  \Pi = \int_{Diff} D\phi\ {\cal U}[\phi] .
  \label{Uintegral}
\end{equation}
Equivalently, we may write the group element as an exponential of 
an algebra element, and formally integrate over the algebra 
rather than over the group, that is
\begin{equation}
  \Pi =  \int D\vec{f}\ e^{-iD[\vec f]}. 
  \label{U}
\end{equation}
This equation has a compelling interpretation as the definition 
of the projector on the kernel of the diffeomorphism constraint 
operator $D_a(x)$ via
\begin{equation}
\Pi \sim \prod_{a,x}\ \delta(D_{a}(x)) \sim \int D\vec{f}\ 
e^{-i\int d^{3}x\ f^{a}(x) D_{a}(x)}
\end{equation}
as in 
\begin{equation}
 \delta(x) = \frac{1}{2\pi}\int_{-\infty}^{+\infty} dp\  e^{-ipx} . 
 \label{delta}
\end{equation}
We shall define the projector on the kernel of the hamiltonian 
constraint in a similar manner. 

\section{A diffeomorphism invariant measure}\label{regularization}

In this section, we construct a measure on the space of scalar 
functions \cite{Higgs}, which will be needed for defining the analog 
of Eq.~{\ref{U}} for the hamiltonian constraint.  Consider smooth 
functions $N: \Sigma \rightarrow S_{1}$ on the three-manifold 
$\Sigma$, taking value on the interval $I=[0, T[$.  We keep track of 
the ``length of the interval $I$'', $T$, instead of normalizing it to 
one, because this will simplify keeping track of dimensions in the 
physical application.  Let $\cal N$ be the space of such functions, 
equipped with the $sup$ topology.  Let $F(N)$ be a continuous complex 
function on the infinite dimensional topological vector space $\cal 
N$, and denote the space of these functions as $\cal L$.  Let 
$\{x_i\}= x_{1}, \ldots, x_{n}$ be a set of (disjoint) points in 
$\Sigma$, and $f:(I)^{n}\rightarrow C$ a complex integrable function 
of $n$ real variables.  Consider a function $F\in{\cal L}$ of the form
\begin{equation}
	F_{\{x_{i}\}, f}(N) = f(N(x_{1}), \ldots ,N(x_{n})), 
\label{gelfand}
\end{equation}
namely a function of $N$ having the the set $\{x_{i}\}$ as its 
domain of dependence.  The set of functions of this form form a 
dense linear subspace of $\cal L$, in the pointwise topology.

The simplest nontrivial of such functions is obtained by picking a 
single point $x$ and choosing $f(N)=N$.  Notice that this defines 
precisely the Gel'fand transform $F_{x}(N)=N(x)$, or in Gel'fand's  
enchanting notation
\begin{equation}
                             x(N) = N(x). 
\label{gt}
\end{equation}
Since the functions of the form (\ref{gelfand}) can be seen as a 
generalization of Gel'fand's $F_{x}(N)$, we denote them as 
``generalized Gel'fand functions'', or simply Gel'fand functions.  
Gel'fand functions can be seen as the scalar-field analog of 
the Ashtekar-Lewandowski's graph-cylindrical functions 
(\ref{al}), which are defined for connection-fields. 

Define the following linear form on the Gel'fand functions
\begin{equation}
   \int DN\  F_{\{x_{i}\}, f}(N) = 
   \frac{1}{(T)^{n}}\int_{I^{n}} dN_{1}\ldots 
    dN_{n}\ f(N_{1}, \ldots , N_{n}). 
\label{integral}
\end{equation}
Here $\frac{dN_{i}}{ T}$ is the normalized invariant measure on 
the interval $I$.  Finally, denote the closure of $\cal L$ in the 
norm
\begin{equation}
	||F|| =  \int DN\  | F(N) |
	\label{L1}
\end{equation}
as $L_{1}[{\cal N}]$; the linear form (\ref{integral}) extends by 
continuity (in the $L^{1}$ topology defined by this norm) to all 
of $L_{1}[{\cal N}]$. 

A simple class of integrable functions is given by polynomial 
Gel'fand functions.  We have indeed
\begin{eqnarray}
	\int DN\ 1 &=& \frac{1}{ T}  \int_{I} dN\ 1  = 1,  
	\nonumber \\
   \int DN\ N(x) &=& \frac{1}{ T} \int_{I} dN  \ N 
			= \frac{1}{ T} \frac{1}{2} T^{2} = \frac{1}{2} T 
	\label{linear}
\end{eqnarray}
Notice that for quadratic functionals we must distinguish two cases
\begin{eqnarray}
\int DN\ N(x)\ N(y) &=& \frac{1}{T^{2}} 
\int_{I}\int_{I} dN_{1}dN_{2}\ 
N_{1}\ N_{2}   = \frac{1}{4}\, T^{2},  \\
\int DN\ N(x)\ N(x) &=& \frac{1}{T} \int_{I} dN\ N^{2}   = 
\frac{1}{ T}\ \frac{1}{3}\, T^{3} = \frac{1}{3}\, T^{2}. 
	\label{quadratic}
\end{eqnarray}
Namely, we must distinguish the case in which the arguments of 
the two functions $N(\ )$ are distinct or the same.  The general 
pattern should be clear.  A general polynomial functional 
$F_{n_{1}\ldots n_{K}}$ will have $n_{k}$ points $x_{1}^{(k)}\ldots 
x_{n_{k}}^{(k)}$ in its domain of dependence
 in which the function $N(x)$ appear with power $k$.  A 
simple calculation yields then
\begin{eqnarray}
I_{n_{1}\ldots n_{K}}  &=& 
\int DN\ F_{n_{1}\ldots n_{K}}  \nonumber \\ &=&
\int DN\ [N(x_{1}^{(1)})\ldots N(x_{n_{1}}^{(1)})] 
[N(x_{1}^{(2)})\ldots N(x^{(2)}_{n_{2}})]^{2} \ldots 
[N(x^{(K)}_{1})\ldots N(x^{(K)}_{n_{K}})]^{K} 
\nonumber \\ 
&=& T^{\sum_{k}n_{k}}\ 
\prod_{k=1}^{K} \left(\frac{1}{k+1}\right)^{n_{k}} \nonumber \\ 
& \equiv & d_{n_{1}\ldots n_{K}}
	\label{d}
\end{eqnarray}

The diffeomorphism group $Diff$ of $\Sigma$ acts 
naturally on $\cal N$, via $(\phi N)(x)=N(\phi^{-1}(x))$, where 
$\phi:\Sigma \rightarrow \Sigma$ is in $Diff$. It is 
easy to see that the integral (\ref{integral}) is 
diffeomorphism invariant
\begin{equation}
	\int [DN]\ F[\phi N] = 	\int [DN]\  F[N]. 
	\label{diffinvariance}
\end{equation}
This follows from the fact that the r.h.s.\ of (\ref{integral}) 
is clearly insensitive to a diffeomorphism transformation on $N$.

\section{Dynamics: the regularized propagator}\label{dynamics}

We now come to the construction of the physical state space $H_{phys}$ 
and the partition function of the theory.  We have to solve the   
Dirac's hamiltonian constraint equation
\begin{equation}
	H[N] \psi = 0 
\end{equation}
for the quantum Hamiltonian constraint $H[N]$. 

\subsection{The Hamiltonian constraint: first version}

The operator $H[N]$ that we consider is a small modification of 
the Riemanian hamiltonian constraint defined in 
\cite{Thiemann}.  However, we make only use of the general 
structure of this operator, which is common to several of the 
proposed variants.  We take a symmetric version of $H[N]$, which 
``creates'' as well as ``destroying'' links.  The matrix elements 
of $H[N]$ are given by
\begin{equation}
\langle \psi |H[N]| \phi \rangle \ = \ 
\langle \psi |C[N]| \phi \rangle \ +\  
\overline{\langle \phi |C[N]| \psi \rangle}
\label{symmetrized}
\end{equation}
where $C[N]$ is the non-symmetric Thiemann's constraint.  

We recall that the operator $C[N]$, acting on a spin network 
state $|s\rangle$, is given by a sum of terms, one per each node 
$i$ of $s$.  Sketchy (a more precise definition will be given  
below), each such term creates an extra link $e_{added}$ 
joining two points, $i'$ and $i'{}'$, on two distinct links 
adjacent to $i$, and alters the colors of the links between $i$ 
and $i'$ and between $i$ and $i'{}'$.  The result is multiplied 
by a coefficient depending only on the colors of $s$, and by the 
value of the smearing function $N$ ``in the point where the node 
$i$ is located''.  This is illustrated in Figure 1.
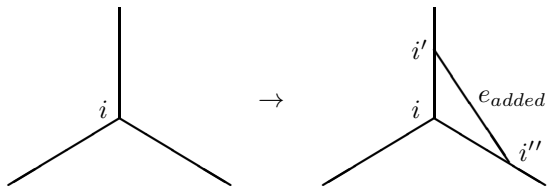
\begin{figure}
\centerline{
\setlength{\unitlength}{.7pt}
\begin{picture}(344,114)(105,606)
\thicklines
\put(201,648){\line(-5,-3){ 60}}
\put(201,648){\line( 5,-3){ 60}}
\put(201,648){\line( 0, 1){ 60}}
\put(371,685){\line( 2,-3){ 40.462}}
\put(371,648){\line(-5,-3){ 60}}
\put(371,648){\line( 5,-3){ 60}}
\put(371,648){\line( 0, 1){ 60}}
\put(271,654){\makebox(0,0)[lb]{\raisebox{0pt}[0pt][0pt]{
$\rightarrow$ }}}
\put(354,680){\makebox(0,0)[lb]{\raisebox{0pt}[0pt][0pt]{
$ i' $}}}
\put(354,648){\makebox(0,0)[lb]{\raisebox{0pt}[0pt][0pt]{
$ i $}}}
\put(185,648){\makebox(0,0)[lb]{\raisebox{0pt}[0pt][0pt]{
$ i $}}}
\put(412,625){\makebox(0,0)[lb]{\raisebox{0pt}[0pt][0pt]{
$ i'{}' $}}}
\put(390,657){\makebox(0,0)[lb]{\raisebox{0pt}[0pt][0pt]{
 $e_{added}$}}}
\end{picture}} 
\caption{Action of the hamiltonian constraint on a trivalent 
node.}
\end{figure}

It is important to observe that $C[N]$ is defined as a map from 
$H_{diff}$ to $H_{aux}^{*}$.  In this definition there is a subtle 
interplay between diff-invariant and non-diff-invariant aspects 
of the hamiltonian constraint, which is a key aspect of the issue 
we are considering, and must be dealt with with care.  The reason 
$C[N]$ is defined on $H_{diff}$, namely on the diffeomorphism 
invariant states, is that it is on these states that the 
``precise position'' of the points $i$ and $i'{}'$ and of the 
link $e_{added}$ is irrelevant.\footnote{More precisely, only on 
these states can the regulator used in the quantization of the 
classical quantity be removed.} However, $C[N]$ is not 
diffeomorphism invariant, and therefore $C[N]|s\rangle$ is not in 
$H_{diff}$, because a diffeomorphism modifies $N$.  The 
feature of $C[N]|s\rangle$ that breaks diffeomorphism invariance 
is the fact that it contains a factor given by the 
value of $N(x)$ in the point in which the node $i$ is located: 
this location is not a diffeomorphism invariant notion.

Before presenting the precise definition of $C[N]$, which takes 
care both of its diff-invariant and its non-diff-invariant 
features, we need to define certain peculiar elements of 
$H_{aux}^{*}$, which will appear in the definition.  Consider an 
s-knot $s$ and let $i$ be one of its nodes.  Let $N$ be a scalar 
function on $\Sigma$.  We define the state $ | s; i, N \rangle$ 
in $H_{aux}^{*}$ by (again, we interchange freely bra and ket 
notation)
\begin{equation}
	\langle s; i, N| S \rangle = 
	N(x_{S,i})\  \langle s | S \rangle. 
\label{defhs}
\end{equation}
where $x_{S,i}$ is the position of the node of $S$ that gets 
identified with the node $i$ of $s$ in the scalar 
product.   Notice that the state $ | s;  i, N \rangle$ is 
``almost'' a diff-invariant s-knot state: in facts, it is 
``almost'' insensible to the location of $S$.  The only aspect of 
this location to which it is sensible is the location of the 
node.  In fact, in the r.h.s.\ of (\ref{defhs}), the diff 
invariant quantity $\langle s | S \rangle$ is multiplied by 
the value of $N(\ )$ {\em in the point in which the node $i$ is 
located}.  The Hamiltonian constraint defined by Thiemann acts on 
diff-invariant states and creates states in $H_{aux}^{*}$, of the 
form (\ref{defhs}).

The precise definition of $C[N]$ is the following
\begin{equation} 
C[N]| s \rangle = A_{i}^{\beta}(s) \ | s_{i}^{\beta}; i, N \rangle. 
\end{equation}
Let us explain our notation.  Sum over the repeated indices 
$\beta$ and $i$ is understood.  The index $i$ runs over the 
nodes in $s$.  The index $\beta=(l',l'{}',\epsilon',\epsilon'{}')$ 
runs over the couples $(l',l'{}')$ of (distinct) links adjacent 
to each node $i$ and over $\epsilon'$ and $\epsilon'{}'$, which 
take the values $+1$ or $-1$.  The s-knot $s_{i}^{\beta}$ was 
introduced in \cite{RSHam}.  It is defined as the right
hand part of Figure 1. That is,  by adding two 
new nodes $i'$ and $i'{}'$ on the two links $l'$ and $l'{}'$ 
(determined by $\beta$) respectively, adding a new link 
$e_{added}$ colored $1/2$ joining $i'$ and $i'{}'$, and altering 
the color of the links joining $i'$ and $i$ (and, respectively, 
$i'{}'$ and $i$) by $+1$ or $-1$ according to the value of 
$\epsilon'$ (respectively $\epsilon'{}'$).  $A_{i}^{\beta}(s)$ 
is a coefficient defined in \cite{Thiemann} whose explicit form 
is computed in \cite{coeff}.  

Summarizing, we have 
\begin{equation}
		\langle S | C[N] | s \rangle = 
		  A_{i}^{\beta}(s) \ 
	N(x_{S,i}) \ \langle S | s_{i}^{\beta} \rangle. 
\end{equation} 
(where $\langle S | s \rangle \equiv \langle s | S 
\rangle=\langle s | s(S) \rangle $; see (\ref{sstate}) and 
(\ref{rl}).)  Clearly, $C[N]$ can equivalently be viewed as an 
operator from $H_{aux}$ to $H_{diff}^{*}$, by writing
\begin{equation}
		\langle s | C[N] | S \rangle = 
		  A_{i}^{\beta}(s) \ 
	N(x_{S,i}) \ \langle s | s(S)_{i}^{\beta} \rangle. 
\label{me}
\end{equation}
(Recall that $s(S)$ is the s-knot to which $S$ belongs.)
This fact allows us to define the symmetrized operator 
(\ref{symmetrized}) by 
\begin{equation}
		\langle S | H[N] | s \rangle = 
		  A_{i}^{\beta}(s) \ 
	N(x_{S,i}) \ \langle S | s_{i}^{\beta} \rangle +
	            \overline{A}_{i}^{\beta}(s) \ 
	N(x_{S,i}) \ \langle s(S)_{i}^{\beta}| s \rangle. 
\end{equation}
Only one of the two terms in the r.h.s.\ of this equation may be 
non-vanishing: the first, if $S$ has two nodes more than $s$; the 
second, if $S$ has two nodes less than $s$.  

We can simplify our notation by introducing an index 
$\alpha=(\beta,\pm 1)$ where $+1$ indicates that a link is added 
and $-1$ indicates that the link is removed. We obtain 
\begin{equation}
		\langle S | H[N] | s \rangle = 
		  A_{i}^{\alpha}(s) \ 
	N(x_{S,i}) \ \langle S | s_{i}^{\alpha} \rangle 
\end{equation}
where 
\begin{equation}
		  A_{i}^{\beta,-1}(s) = 
		 \overline A_{i}^{\beta,+1}(s_{i}^{\beta}). 
\end{equation}

Notice that the precise position of $e_{added}$ (and thus $i'$ and 
$i'{}'$) drops out from the final formula, because of the diff 
invariance of the quantity $\langle S | s_{i}^{\alpha} \rangle 
$.\footnote{One must only worry about the positioning of $e_{added}$ 
up to isotopy.  This is carefully defined in \cite{Thiemann}, 
following a construction by Lewandowski.} This is essential, because 
if a specific position for $e_{added}$ had to be chosen, 
diffeomorphism<<<<<<w invariance would be badly broken.  One can view the 
coordinate distance $\epsilon$ between $i$ and $i'$ (and between $i$ 
and $i'{}'$) as a regulating parameter to be taken to zero {\em 
after\/} the matrix elements (\ref{me}) have been evaluated.  The 
limit $\epsilon\to 0$ is discontinuous, but $\langle s | 
S_{i}^{\beta}\rangle$ is independent from $\epsilon$ for $\epsilon$ 
sufficiently small, and therefore the limit of these matrix elements 
is trivial.  Thus, the operator is defined thanks to two key tricks:
\begin{enumerate}  
\item  the diff invariance of the state $|s\rangle$ acted upon 
allows us to get rid of the precise position of $e_{added}$; 
\item   the {\em  lack\/ } of diff invariance of $\langle S|$, 
allows us to give meaning to the point $x_{S,i}$ ``where 
the node is located'', and therefore allows us to give meaning to 
the smearing of the operator with a given (non diff invariant!)  
function $N(x)$.  
\end{enumerate} 
This is why $H[N]$ is defined as a map from $H_{diff}$ to 
$H_{aux}^{*}$.  The first of these two key facts, which allow the 
quantum hamiltonian constraint operator to exist, was recognized in 
\cite{RSHam}, the second in \cite{Thiemann}. 

\subsection{The Hamiltonian constraint: second version}
\label{products}

The interplay between diff invariant and non diff invariant 
constructs described above needs to be crafted even more finely, 
in order to be able to exponentiate the Hamiltonian constraint 
and derive its kernel.  In fact, in order to exponentiate $H[N]$ 
and to expand the exponential in powers, we will have to deal 
with products of $H[N]$'s.  In order for these products to be 
well defined, the domain of the operator must include its range, 
which is (essentially) $H_{aux}$.  Therefore we need to extend 
the action of $H[N]$ from $H_{diff}$ to $H_{aux}$.  The price for 
this extension is, of course, that the operator becomes dependent 
on the regulator, namely on the precise position in which 
$e_{added}$ is added.  However, we can do so {\em here\/}, 
because such dependence will disappear in the integration over 
$N$ !

We define $H[N]$ on $H_{aux}$ by simply picking a particular 
position for $e_{added}$ in the definition of $S_{i}^\alpha$
\begin{equation}
  H[N]| S \rangle = N(x_{i})\  A_{i}^{\alpha}(S) \ 
|S_{i}^{\alpha}\rangle. 
\end{equation}
where, clearly, $A_{i}^{\alpha}(S)= A_{i}^{\alpha}(s(S))$. (When 
$\alpha=(\beta,-1)$, no modification is necessary.  That is, a 
link is removed irrespectively from its precise location.)   
In a quadratic expression this yields 
\begin{equation}
	 \langle S | H[N] H[N] | s \rangle = 
	 A_{i}^{\alpha}(S_{i_{1}}^{\alpha_{1}}) \ 
	 A_{i}^{\alpha}(S) \ 
	 N(x_{S{}_{i_{1}}^{\alpha_{1}}{}_{i_{2}}^{\alpha_{2}},i_{2}}) 
	 \ N(x_{S_{i_{1}}^{\alpha_{1}},i_1}) \ \langle S | 
	 s{}_{i_{1}}^{\alpha_{1}}{}_{i_2}^{\alpha_{2}} \rangle.
\label{quadratic2}
\end{equation}
Here $i_{1}$ labels the nodes of $s$.  After the action of the 
first operator, and thus the addition (or subtraction) 
of one link, we obtain 
$s_{i_{1}}^{\alpha_{1}}$.  The index $i_{2}$ runs over the nodes 
of $s_{i_{1}}^{\alpha_{1}}$, and therefore its range is larger 
(or smaller) than the index $i_{1}$, because of the two new 
nodes (or the two nodes removed).  

Notice that in each of the terms of the sum in the r.h.s.\ of 
(\ref{quadratic2}) (that is, for each fixed value of the indices 
$i_{1}, \alpha_{1}, i_{2}, \alpha_{2}$) we have a product
\begin{equation}
	N(y)\ N(x)
\label{pippo}  
\end{equation} 
where $x$ and $y$ are the positions of the two nodes acted upon by the 
two operators.  In particular, the second $H[N]$ operator may act on 
one of the nodes created by the first operator $H[N]$.  For instance, 
$x$ in (\ref{pippo}) may be the coordinates of the point $i$ in (the 
l.h.s.\ of) Figure (1), and $y$ may be the coordinates of the point 
$i'$ (in the r.h.s.\ of the Figure).  Now, later on, expressions such 
as (\ref{quadratic2}) will appear within functional integrals over 
$N$.  Inside these integrals, the only feature of these two positions 
that matter is whether $x=y$ or not (see section 
\ref{regularization}).  Therefore, the only feature of the position of 
$e_{added}$ that matters is whether its end points, namely $i'$ and 
$i'{}'$ in Figure 1, are on top of $i$ or not.  This is the only 
dependence on the regulator (the position of $e_{added}$) that 
survives in the integral !  More precisely, in the integration over 
$N$, the arbitrariness in the regularization reduced to the 
arbitrariness of the decision of whether or not we should think at 
$i$, $i'$ and $i'{}'$ in the r.h.s.\ of Figure 1, as on top of the 
point $i$ on the l.h.s.\ or not.

We can view this choice in the following terms.  The operator $C[N]$ 
creates new nodes at positions which are displaced from the original 
node by a distance $\epsilon$, where $\epsilon$ is to be later taken 
to zero (taking this limit is in fact necessary in order to identify 
the quantum operator with the desired classical quantity).  The choice 
is whether to take $\epsilon$ to zero before or after the integration 
over $N$.

Let us denote the position of the node $i$ of $S$ (the node acted 
upon) by $x$.  Denote the two new nodes created by the action of 
the operator as $y'$ and $y'{}'$.  And denote the position of the 
node $i$ after the action of the operator as $y$ (nothing forces 
$x=y$ a priori).  The natural choices are
\begin{enumerate}  
\item $y=x$, $y'\ne x$, $y'{}' \ne x$, 
\item $y \ne x$, $y'\ne x$, $y'{}'\ne x$, 
\item $y=x$, $y' = x$, $y'{}' = x$. 
\end{enumerate}
The choice is exquisitely quantum field theoretical: we are 
defining here the product of operator valued distributions, and 
we encounter an ambiguity in the renormalization of the 
regularized product.  We thus have three options for the 
regularization of the operator products $C[N] \ldots C[N]$, 
corresponding to the three choices above.  

Choice 3 is not (easily) compatible with the symmetrization of the 
operator, and choice 2 yields a nonsensical vanishing of all the 
matrix elements of the projector.  Thus, we adopt, at least 
provisionally, choice 1 (which, after all, is probably the most 
natural).  That is, we assume that $i$ itself is not displaced by the 
hamiltonian constraint operator, while $i'$ and $i'{}'$ are created in 
positions which are {\em distinct\/} from the position of $i$ (see 
Figure 1).

For a product of $n$ operators (with the same smearing function), 
we have 
\begin{equation}
	 \langle S | (H[N])^{n} | s \rangle  =    
	 N(x_1)  \ldots N(x_n) \ 
	 A_{i_{1}}^{\alpha_{1}}(S) \ldots 
	 A_{i_{n}}^{\alpha_{n}}(S_{i_{1}\ldots 
	 i_{n-1}}^{\alpha_{1}\ldots \alpha_{n-1}}) \ 
	  \langle S | {s'}_{i_{1}\ldots 
	 i_{n}}^{\alpha_{1}\ldots \alpha_{n}} \rangle,
\label{power}
\end{equation} 
where $i_{j}$ runs over the nodes of ${s'}_{i_{1}\ldots 
i_{j-1}}^{\alpha_{1}\ldots \alpha_{j-1}}$, and we have denoted simply 
as $x_{1}\ldots x_{n}$ the positions of the sequence of nodes acted 
upon in a given term.  According to the regularization chosen, this 
sequence contains points which are distinct except when a node is 
acted upon repeatedly. 

\subsection{Expansion}

Our task is now to define the space $H_{phys}$, using the various 
tools developed above.  We aim at defining $H_{phys}$ following the 
lines $H_{diff}$ is defined by the operator $\Pi$ in equations 
(\ref{U}) and (\ref{rl}).  That is, we want to construct the operator
\begin{equation}
  P =  \int DN \ e^{-iH[N]}, 
\label{P}
\end{equation}
whose matrix elements
\begin{equation}
\langle s | P | s' \rangle 
= \int DN \  \langle s | e^{-iH[N]} | s' \rangle. 
\label{sPsH}
\end{equation} 
define the quadratic form 
\begin{equation}
\langle s | s' \rangle_{phys} = \langle s | P | s' \rangle.  
\end{equation}
$H_{phys}$ is then the Hilbert space defined over the pre-Hilbert 
space $H_{diff}$ by the quadratic form $\langle \ |\ \rangle_{phys}$.  
As for $\Pi$ (see section \ref{diff}), we will call $P$ a 
``projector'', slightly forcing the usual mathematical meaning of this 
term.

Notice that the Hamiltonian constraint we use is a density of 
weight one (instead than two, as in the original Ashtekar 
formalism); therefore the integration variable $N$ is a scalar 
field.  This fact will allow us to interpret the integration in 
$N$ in terms of the integral defined in section 
\ref{regularization}.\footnote{%
One might be puzzled by the fact that the measure defined in section 
\ref{regularization} is normalized, while the measure in equation 
(\ref{delta}), which is the formal analog of the expression (\ref{P}), 
must not be normalized, nor can be seen as the limit of normalized 
measures.  The problem, however, is that the choice of the measure in 
(\ref{P}) must incorporate the renormalization of the divergence 
coming (at least) from the volume of the gauge orbit.  The 
normalization of the measure is needed to make our expressions 
converge, and should be viewed, we think, as a quantum field 
theoretical subtraction.} The importance of having a weight-one 
hamiltonian constraint in the quantum theory, was realized by Thiemann 
\cite{Thiemann}.

We begin by regularizing the integral (\ref{sPsH}) by restricting 
the integration domain of the functional integral in $[DN]$ to 
the subdomain formed by all the functions $N$ that satisfy
\begin{equation}
	|N(x)| < T 
	\label{T} 
\end{equation} 
where $T$ is a regularization parameter with the dimensions of a 
time.  The physical limit is recovered for $T\to\infty$.  We write
\begin{equation}
\langle s | P_{T} | s' \rangle 
\equiv \int_{|N(x)|<T}  DN \  \langle s | e^{-iH[N]} | s' \rangle.   
\label{regT}
\end{equation} 
Notice that the regularization (\ref{T}) is diffeomorphism 
invariant. 

By taking advantage from the diff invariance of the expression 
(\ref{regT}), we can insert an integration over the
diffeomorphisms and rewrite (\ref{regT}) using (\ref{diffsmear}) as
\begin{equation}
\langle s | P_{T} | s' \rangle 
= \int_{Diff} D\phi \int_{|N(x)|<T}  DN \  
\langle  {\cal U}(\phi)S | e^{-iH[N]} | s' \rangle  
\end{equation} 
where $S$ is any spin network such that 
\begin{equation}
s(S)=s. 	
	\label{sS}
\end{equation}
Next, we expand the exponent in powers
\begin{equation}
\langle s | P_{T} | s' \rangle 
= \int_{Diff} D\phi \int _{|N(x)|<T} DN\ \langle {\cal U}(\phi) S | 
\left( 
\sum_{n=0}^{\infty} \frac{(-i)^{n}}{n!} (H[N])^{n} \right) 
 | s' \rangle. 
\end{equation} 
Using the explicit form (\ref{power}) of the hamiltonian 
constraint operator and acting with ${\cal U}(\phi)$ explicitly 
we obtain
\begin{eqnarray}
\langle s | P_{T} | s' \rangle 
&=& \sum_{n=0}^{\infty} \frac{(-i)^n}{n!}
\int_{Diff} D\phi \int_{|N(x)|<T}  DN\ 	
N(\phi(x_{i_{n}})) \ldots N(\phi(x_{i_{1}})) \nonumber \\ 
&& \times 
A_{i_{n}}^{\alpha_{n}}(s_{i_{1}\ldots i_{n-1}}^{\alpha_{1}\ldots 
\alpha_{n-1}}) \ldots A_{i_{1}}^{\alpha_{1}}(s) \ 
\langle s | {s'}_{i_{1}\ldots i_{1}}^{\alpha_{1}\ldots \alpha_{n}} 
\rangle,
\end{eqnarray} 
where $i_{j}$ runs over the nodes of ${s'}_{i_{1}\ldots 
i_{j-1}}^{\alpha_{1}\ldots \alpha_{j-1}}$. We have also used
(\ref{sS}) and 
\begin{equation}
	s(S_{i_{1}\ldots i_{j}}^{\alpha_{1}\ldots\alpha_{j}})=	
	s_{i_{1}\ldots i_{j}}^{\alpha_{1}\ldots\alpha_{j}}. 
\end{equation}
which follows from it.   Notice that, as 
promised, the only remaining diff-dependent quantities are the arguments 
of the functions $N(\ )$.  But since the $DN$ integral is diff invariant (see 
Eq.~(\ref{diffinvariance})), the integration over $Diff$ 
can be trivially performed using (\ref{reg1}).   Also, 
notice that  the $N(x)$'s appear only in the polynomials.  
Thus we have
\begin{equation}
\langle s | P_{T} | s' \rangle 
= \sum_{n=0}^{\infty}\ \frac{(-i)^n}{n!}\ 
I_{x_{i_{n}} \ldots x_{i_{1}}}(T) \ \ 
A_{i_{n}}^{\alpha_{n}}(S_{i_{1}\ldots i_{n-1}}^{\alpha_{1}\ldots 
\alpha_{n-1}}) \ldots A_{i_{1}}^{\alpha_{1}}(S) \ 
\langle s | {s'}_{i_{1}\ldots i_{1}}^{\alpha_{1}\ldots \alpha_{n}}  
\rangle,
\label{eccolo}
\end{equation} 
where 
\begin{equation}
I_{x_{i_{n}} \ldots x_{i_{1}}}(T) = \int_{|N(x)|<T} DN\ 
N(x_{i_{n}}) \ldots N(x_{i_{1}}). 
\end{equation}

Now, the last integral is precisely the integral of a 
polynomial Gel'fand function discussed in the previous section. 
The only difference here is that the domain of the $dN$ integral 
is between $-T$ and $T$ instead than between $0$ and $T$.  The 
effect of this is just to put all the odd terms to zero and to 
double the even terms.  Let $n_{k}$ be the number of points that 
appear $k$ times in the list $x_{i_{n}} \ldots x_{i_{1}}$, so 
that
\begin{equation}
\sum_{k} k n_{k} = n. 
\label{f}
\end{equation}
We obtain
\begin{equation}
     I_{x_{i_{n}} \ldots x_{i_{1}}}(T) = 
     \left( \prod_{k} e(k) \right) d_{n_{1}\ldots n_{k}}
\end{equation}
where $d_{n_{1}\ldots n_{k}}$ is defined in Eq.~(\ref{d}), and 
$e(k)$ is defined for any integer $n$ by 
\begin{eqnarray}
e(2n) & = & 2, \nonumber \\ 
e(2n+1) & = & 0. 
\end{eqnarray} 
From (\ref{d}), and (\ref{f}), we have 
\begin{equation}
     I_{x_{i_{n}} \ldots x_{i_1}}(T) = 
     \prod_k e(k) \left(\frac{T^k}{k+1}\right)^{n_k}=
     T^n \prod_k \frac{e(k)}{(k+1)^{n_k}} 
\label{ff}
\end{equation}
Inserting (\ref{ff}), in (\ref{eccolo}), we conclude
\begin{eqnarray}
\langle s | P_{T} | s' \rangle 
&=& \sum_{n=0}^{\infty}\ \   T^{n} \ \  
\langle s | P^{(n)}| s' \rangle   \label{PT}
\label{final1} \\
\langle s | P^{(n)}| s' \rangle & = & \frac{(-i)^{n}}{n!}  
\sum_{i_{1}\ldots i_{n}\alpha_{1}\ldots 
\alpha_{n}} 
\prod_{k} \ \frac{e(k)}{(k+1)^{n_k}} \ \ 
A_{i_n}^{\alpha_n}(s_{i_1\ldots i_{n-1}}^{\alpha_1\ldots 
\alpha_{n-1}}) \ldots A_{i_1}^{\alpha_1}(s) \ \langle s | 
{s'}_{i_1\ldots i_1}^{\alpha_1\ldots \alpha_n} \rangle
\label{final}
\end{eqnarray}
We recall that the technique for the explicit computation of the 
coefficients $A_{i}^{\alpha}(s) $ is given in \cite{coeff}.  The last 
equation is an explicit and computable expression, term by term 
finite, for the regularized matrix elements of the projector on the 
physical state space of the solutions of the hamiltonian constraint.

\subsection{Interpretation: spin foam}

The terms of the sum (\ref{final}) are naturally labeled by 
branched colored surfaces \cite{rr,Baez,Reisenberger},  or ``spin 
foams''.  Each 
surface represents a history of the s-knot state.  More 
precisely, consider a finite sequence $\sigma_{n}$ of $n+1$ 
spin networks
\begin{equation}
s_{0}, s_{1} \ldots,  s_{n}
\label{sequence}
\end{equation}
In particular, let the sequence (\ref{sequence}) be 
generated by a sequence of $n$ actions of single terms of 
the Hamiltonian constraint acting on $s_{0}$
\begin{equation}
\sigma_{n} = \left\{s,\ s_{i_{1}}^{\alpha_{1}}, \ 
s_{i_{1}i_{2}}^{\alpha_{1}\alpha_{2}},\ \ldots ,\  
s_{i_{1}\ldots i_{n}}^{\alpha_{1}\ldots \alpha_{n}} 
\right\} 
\label{sequence2}
\end{equation} 
We call such a sequence a ``spin foam'', and we represent it as a 
branched colored 2d surface.  A branched colored surface is a 
collection of elementary surfaces (faces) carrying a color.  The 
faces join in edges carrying an intertwiner.  The edges, in turn, 
join in vertices.  A branched colored surface with $n$ vertices 
can be identified with a sequence (\ref{sequence2}) if it can be 
sliced (in ``constant time'' slices) such that any slice that 
does not cut a vertex is one of the spin networks in 
(\ref{sequence2}).  In other words, the branched colored surface 
can be seen as the spacetime world-sheet, or world-history of the 
spin network that evolves under $n$ actions of the hamiltonian 
constraint.  

Each action of the hamiltonian constraint splits a node of the 
spin network into three nodes (or combine three nodes into one), 
and thus generates a vertex of the branched surface.  Thus, as in 
the usual Feynman diagrams, the vertices describe the elementary 
interactions of the theory.  In particular, here one sees that 
the complicated action of the hamiltonian displayed in Figure 1, 
which makes a node split into three nodes, corresponds to the 
simplest geometric vertex.  Figure 2 is a picture of the 
elementary vertex.  Notice that it represents nothing but the 
spacetime evolution of the elementary action of the hamiltonian 
constraint, given in Figure 1.
\begin{figure} 
\centerline{\mbox{\epsfig{file=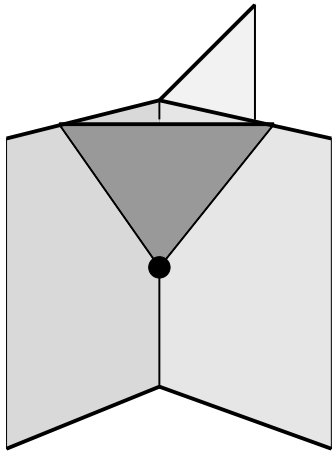}}} 
\caption{The elementary vertex.} 
\end{figure}
An example of a surface in the sum is given in Figure 3.
\begin{figure} \centerline{\mbox{\epsfig{file=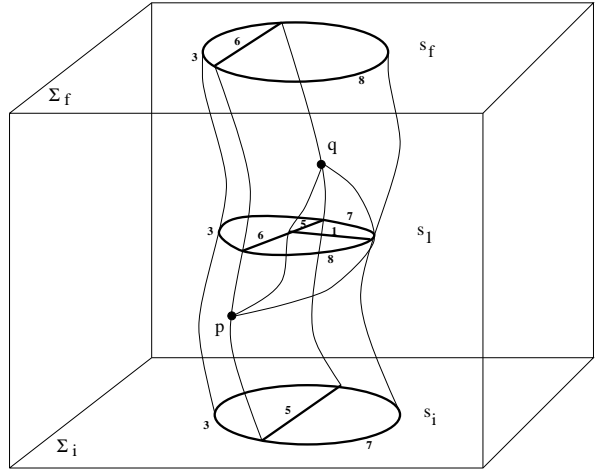}}}
 \caption{A term of second order.}
\end{figure}
We write $\partial\sigma_{n}=s\cup s'$ to indicate that the spin foam 
$\sigma_{n}$ is bounded by the initial and final spin networks $s$ and 
$s'$.  We associate to each $\sigma_{n}$ the amplitude
\begin{equation}
A(\sigma_{n}) =  \frac{1}{n!}  \prod_{v=1}^{n} A(v)
\prod_{k} \frac{e(k)}{(k+1)^{n_{k}}} 
\label{asigma}
\end{equation}
where $v$ run over the vertices of $\sigma_{n}$ and the 
amplitude of a single vertex is
\begin{equation}
A(v) = A_{i_{v}}^{\alpha_{v}}(s_{i_{1}\ldots i_{v-1}}^{\alpha_{1}\ldots 
\alpha_{v-1}}). 
\label{Av}
\end{equation}
The amplitude of a vertex depends only on the coloring of 
the faces and edges adjacent to the vertex. 

Using this, we can rewrite Eqs.~(\ref{PT}, \ref{final}) as 
\begin{equation}
\langle s | P_{T} | s' \rangle = \sum_{n=0}^{\infty}\ \  T^{n} 
\sum_{\sigma_{n}, \partial\sigma_{n}=s\cup s'} 
A(\sigma_{n}).
\label{ss}
\end{equation}

The key novelty with respect to \cite{rr} is the factor 
\begin{equation}
           \prod_{k} \frac{e(k)}{(k+1)^{n_{k}}} 
\end{equation}
The integers $n_{k}$ are determined by the number of multiple actions 
of $H[N]$ on the same vertex.

The last expression leads immediately to the form of the (regularized) 
``vacuum to vacuum'' transition amplitude, or the partition function 
of the theory
\begin{equation}
Z_{T} = \sum_{n=0}^{\infty} T^{n}\ \sum_{\sigma_{n}}  A(\sigma_{n})
\label{Z}
\end{equation} 
for $\sigma$'s with no boundaries.  In words, the theory is defined as 
a sum over spin foams $sigma_{n}$, where the amplitude $A(\sigma_{n})$ 
of a spin foam is determined, via (\ref{asigma}), by the product of 
the amplitudes $A(v)$ of its vertices.  Thus, the theory is determined 
by giving the amplitude $A(v)$ of the vertex, as a function of 
adjacent colors.

\subsection{Physical observables}

The expressions we have defined depend on the regulator $T$.  A 
naive limit $T\to\infty$ yields to meaningless divergences.  On 
the other hand, it is natural to expect to be able to remove the 
regulator only within expressions for physical expectation 
values.  Loosely speaking, the integral (\ref{P}) defines a 
delta-like distribution, and does not converge to any 
function itself; however, its contraction with a smooth function 
should converge.  In particular, for finite $T$ the integral 
contracted with a function gives the integral of the Fourier 
transform of the function over the interval $[-T, T]$.  If the 
function has a Fourier transform that decays reasonably fast, 
then the integral should converges nicely.  Thus, we may expect 
the expansion in $T$ to be meaningful for suitable observables 
(see next section).

The difficulty of constructing interesting 
physical observables invariant under four-dimensional 
diffeomorphisms in general relativity in well known 
\cite{observables} and we do not discuss this problem here.  
Instead, we notice that given an operator $A$ on $H_{diff}$, 
invariant under three-dimensional diffeomorphisms, one can 
immediately construct a fully gauge invariant operator $O$ simply 
by
\begin{equation}
	O = P\ A\ P. 
	\label{0}
\end{equation}
For instance, $A$ may be the volume $V$ of $\Sigma$ operator 
\cite{discreteness,volume}; or the projector on 
a given eigenspace of $V$
 \begin{equation}
 	A = \delta(V,v)
 \label{Avv}
 \end{equation}
where $v$ is one of the eigenvalues of $V$.  
Consider the 
expectation value of $O$ in a physical state
\begin{equation}
	\langle O \rangle  = \frac{\langle s | O | s \rangle_{phys} 
	} {\langle s | s \rangle_{phys} } = 
	\frac {\langle s |  P A P | s \rangle } {\langle s |P| s \rangle }. 
	\label{expectation}
\end{equation} 
While we expect this quantity to be finite (for an appropriate 
$A$), the numerator and the denominator are presumably 
independently divergent, as one may expect in a field theory.  
Our strategy to compute $\langle O \rangle$, therefore, must be 
to take the $T\to\infty$ limit of the ratio, and not of the 
numerator and of the denominator independently.  We thus properly 
define 
\begin{equation}
\langle O_{T} \rangle = 
\frac{\langle s |  P_{T} A P_{T} | s \rangle } {\langle s P_{T} s 
\rangle }. 
\label{ot}
\end{equation}
and 
\begin{equation}
\langle O \rangle=\lim_{T\to\infty} \langle O_{T} \rangle.
\label{Tt}
\end{equation}

Both the numerator and the numerator in (\ref{ot}) can be written 
as power series in $T$.  Therefore we have 
\begin{equation} 
\langle O \rangle  = \lim_{T\to\infty} 
\frac{\sum_{n}T^{n} a_{n}}{\sum_{m}T^{m}b_{m}}
\label{ratio}
\end{equation}
where
\begin{equation}
a_{n} =
\sum_{m=0,n}
\langle s | P^{(m)}_{T} | s' \rangle \ 
\langle s' | V | s'' \rangle \ 
\langle s'' | P^{(n-m)}_{T} | s' \rangle 
\label{an}
\end{equation}
and 
\begin{equation}
b_{n} = \langle s | P^{(n)}_{T} | s \rangle.  
\label{bn} 
\end{equation}
The matrix elements $\langle s | P^{(n)}_{T} | s \rangle$ are 
explicitly given in (\ref{final}).  Notice that they are finite 
and explicitly computable.  Equation (\ref{ot}) defines a 
function of $T$ analytic in the origin.  We leave the problem of 
determining the conditions under which the higher order terms are 
small, and of finding techniques for analytically continuing it 
to infinity on the Riemann sphere, for future investigations.

\subsection{Quantum ADM surfaces}

An important lesson is obtained by writing the expression for the 
expectation values in the spin foam version.  Consider, for 
simplicity, the case in which $A$ is diagonal in the loop basis.  
(This is true for the volume, which is the reason for the choice of 
the basis in the intertwiners space, in section \ref{loop}).  In this 
case, (\ref{an}) becomes
\begin{equation}
a_{n} = 
\sum_{m=0,n}
\langle s | P^{(m)}_{T} | s' \rangle \ 
A(s') \langle s' | P^{(n-m)}_{T} | s \rangle .
\end{equation}
Recalling (\ref{final}) and (\ref{ss}), this can be rewritten as 
\begin{equation}
a_{n} =  \sum_{\sigma_{n}} \left(\sum_{s'} V(s')\right)  
A(\sigma_{n}) 
\label{cuts}
\end{equation}
where the $s'$ are all possible spin networks that cut the surface 
$\sigma_{n}$ into two (past and future) parts.  

Equation (\ref{cuts}) shows that the expectation value of $O$ is 
given by the average of $A$ on all the {\em discrete 
ADM-like spatial slices \/}   $s'$ that cut the quantum spin foam. 
Summarizing,  
\begin{equation}
\langle O \rangle  = \lim_{T\to\infty} 
\frac{\sum_{n}T^{n} \sum_{\sigma_{n}} \big(\sum_{s'} A(s')\big)  
A(\sigma_{n}) }{\sum_{m}T^{m}\sum_{\sigma_{m}}  A(\sigma_{m})}; 
\label{cuts2}
\end{equation}
The sum in $\sigma_{n}$ is over spin foams (with $n$ vertices).  
For every spin foam, the sum in $s'$ is over all its 
``spacelike'' slices.   

This is a nice geometric result.  It clarifies the physical 
interpretation of the four-dimensional space generated by the 
expansion in $T$: it is the quantum version of the four-dimensional 
spacetime of the classical theory.  To see this, consider for instance 
the observable defined in (\ref{Avv}), namely the projector on a given 
eigenspace of the volume.  Classically, the volume is defined if a 
gauge-fixing that identifies a spacelike ``ADM'' surface $\Sigma$ is 
given.  In (\ref{cuts2}), we see that in the quantum theory the role 
of this spacelike surface is taken by the ``ADM-like'' spatial slices 
of the quantum spin foam $s'$.  Thus, we must identify the surfaces 
$s'$ on the spin foam with the classical ADM-surfaces (both are gauge 
constructs!), and therefore we must identify the spin foam itself as 
the (quantum version of the) four-dimensional spacetime of the 
classical theory.\footnote{I thank Mike Reisenberger for this 
observation.}

If the spin foam represents spacetime, the expansion parameter 
$T$ --introduced above simply as a mathematical trick for 
representing the delta function as the integral of an 
exponential-- can be identified as a genuine time variable (it 
has the right dimensions).  This fact provides us with an 
intuitive grasping on the regime of validity of the expansion 
itself.  It is natural to expect that (\ref{Tt}) might converge 
for observables that are sufficiently ``localized in time''.  
These are precisely the relevant observables in the classical 
theory as well.  We illustrate them in the following section. 

\subsection{4d diff-invariant observables can be localized in time}

Claims that diffeomorphism invariant observables cannot be localized 
in time can be found in the literature, and have generated much 
confusion.  These claims are mistaken.  Let us illustrate why a 
physical general relativistic measurement localized in time is 
nevertheless represented by a diffeomorphism invariant quantity.

Consider a state of the solar system.  The state can be given by 
giving positions and velocities of the planets and the value of the 
gravitational field on a certain initial ADM-surface -- or, 
equivalently, at a certain coordinate time $t$.  We can ask the 
following question: ``How high will Venus be on the horizon, seen from 
Alexandria, Egipt, on sunrise of Ptolemy's 40th birthday?''  In 
principle, this quantity can be computed as follows.  First, solve the 
Einstein equations by evolving the initial data in the coordinate time 
$t$.  This can be done using an arbitrary time-coordinate choice, and 
provides the Venus horizon height $h(t)$.  Then, search on the 
solution for the coordinate time $t_{Pt}$ corresponding to the 
physical event used to specify the time (sunrise time of Ptolemy's 
birthday, in the example).  The desired number is finally 
$H=h(t_{Pt})$, which is a genuine diff-invariant observable, 
independent from the coordinate time $t$ used.  The quantity $H$ is 
coordinate-time independent, but it is also well localized in time.

In practice, there is no need for computing the solution of the 
equations of motion {\em for all times $t$.} It is sufficient to 
evolve just from $t$ to $t_{Pt}$, and if $t$ and $t_{Pt}$ are 
sufficiently close, an expansion in $(t_{Pt}-t)$ can be 
effective.  

The same should happen in the quantum theory.  We characterize the 
state of the solar system by means of the (non-gauge-invariant) state 
$|s\rangle$, describing the system at a coordinate time $t$.  If we 
are interested in the expectation value of an observable $H$ at a time 
$t_{Pt}$ and if $t_{Pt}$ is sufficiently close to $t$, we may then 
expect, on physical grounds, that the expansion (\ref{cuts2}) be well 
behaved.  In other words, we do not need to evolve the spin network 
state $|s\rangle$ forever, if what we want to know is something that 
happens shortly after the moment in which the quantum state is 
$|s\rangle$.   Nevertheless, what we are computing is 4-d diff 
invariant quantity.

As an example that is more likely to be treatable in the quantum 
theory, consider an observable such as the volume $V_{c}$ of the 
constant-extrinsic-curvature ADM slice with given extrinsic curvature 
$K(x)=k$.  Assume we have an operator $K$ corresponding to the local 
extrinsic curvature.  Then, 
\begin{equation}
	 V_{k} \sim   P\  \delta(K,k)\ V  \ \delta(K,k)\ P. 
\label{Vc}
\end{equation}
Once more, $V_{c}$ is a 4d-diffeomorphism invariant observable 
localized in time.  Given an extrinsic curvature operator $K$, 
the methods developed here should provide an expansion for the 
expectation value of $V_{k}$.  Inserting (\ref{Vc}) in 
(\ref{cuts2}), with $O=V_{k}$, the delta function selects the ADM 
slices with the correct extrinsic curvature from the sum in $s'$, 
and the mean value of $V_{k}$ is given by the average over such 
slices appearing in the time development of $|s\rangle$ 
generated by the Hamiltonian constraint.  If $|s\rangle$ is 
sufficiently close (in time) to a $K=k$ surface, then, on 
physical grounds, we have some reasons to hope that the expansion 
$T$ to be well behaved.  Thus, the diffeomorphism invariant 
quantum computation of $V_{k}$ reproduces the structure of the 
classical computation.

\section{Conclusions} 

We have studied the dynamics of nonperturbative quantum gravity.  
Because of the diffeomorphism invariance of the theory, this dynamics 
is captured by the ``projector'' $P$ on the physical states that solve 
the hamiltonian constraint.  We have constructed an expansion for the 
(regularized) projector $P$, and for the expectation values of 
physical observables.  The expansion is constructed using some formal 
manipulations and by using a diffeomorphism invariant functional 
integration on a space a scalar functions.  This construction may 
represent a tool for exploring the physics defined by various 
hamiltonian constraints.

Our main result is summarized in equations 
(\ref{final1}-\ref{final}), which give the regularized projector 
and equations (\ref{ratio}-\ref{bn}), which gives the expectation 
value of a physical observable, both in terms of finite and 
explicitly computable quantities.  Equivalently, the theory is 
defined in the spin-foam formalism by the partition function $Z$ 
given in equations (\ref{asigma}-\ref{Av}-\ref{Z}).  The 
expectation values are then given in equation (\ref{cuts2}) as 
averages over the spin foam.  The spin-foam formalism is 
particularly interesting, because it provides a spacetime 
covariant formulation of a diffeomorphism-invariant theory.  The 
partition function $Z$ is expressed ``\`a la Feynman'' as a sum 
over paths, but these paths are {\em topologically\/} distinct, 
and discrete (so that we have a sum rather than an integral).  

Several aspects of our construction are incomplete and deserve more 
detailed investigations.  (i) The physical discussion on the range of 
validity of the expansions considered should certainly be made more 
mathematically precise.  (ii) The choice of the position of the nodes 
considered in Section \ref{products} is somewhat arbitrary and other 
possibilities might be explored.  (iii) A possible modification of the 
formalism that could be explored is to restrict the range of the 
integration in $N$ to positive definite $N$'s, in analogy with the 
Feynman propagator.  (iv) The spacetime geometry of the (individual) 
spin foams has not yet been fully understood and deserves extensive 
investigations (on this, see 
\cite{Baez,Reisenberger,Reisenberg97,geometria,Barbieri}).  (v) We 
have completely disregarded the Lorentzian aspects of the theory.  
These can be taken into account by using the Barbero-Thiemann 
Lorentzian hamiltonian constraint \cite{Thiemann}, or, alternatively, 
along the lines explored by Smolin and Markopoulou \cite{Marko}.  (vi) 
The limit $T \to\infty$ should be better understood: how can we find 
it from the knowledge of a finite number of the $a_{n}$ and $b_{n}$ 
coefficients in (\ref{ratio})?  (vii) Finally, the formalism developed 
here makes contact with the spin foam models 
\cite{BarretCrane,Reisenberg97,geometria}) We believe that the 
relation between these approaches deserves to be studied in detail.

The key issue is whether the measure that we have emploied is the 
``correct'' one.  Intuitively, whether this measure has the property 
that the integral of the exponential gives the delta function, or 
whether $P$ is in fact, in the appropriate sense, a projector.  We 
will discuss this point elsewhere.

The nonperturbative dynamics of a diffeomorphism invariant quantum 
field theory is still a very little explored territory; the scheme 
proposed here might provide a path into this unfamiliar terrain.

\vskip.5cm

I thank Don Marolf, Mike Reisenberger, Roberto DePietri, Thomas 
Thiemann and Andrea Barbieri for important exchanges and for numerous 
essential clarifications.  This work was supported by NSF Grant 
PHY-95-15506, and by the Unit\`e Propre de Recherche du CNRS 7061.

\end{document}